\documentclass[10pt]{article}
\usepackage[OE]{express}
\usepackage{graphicx} % support the \includegraphics command and options
\usepackage{subcaption}
\usepackage{float}
\usepackage{mathrsfs,amsmath}   %The amsmath package is included for \xrightarrow

\begin{document}

\title{Optimized phase sensing in a truncated SU(1,1) interferometer}

\author{Prasoon Gupta,\authormark{1} Bonnie L. Schmittberger,\authormark{1} Brian E. Anderson,\authormark{2} Kevin M. Jones,\authormark{3} and Paul D. Lett\authormark{1,4,*}}

\address{\authormark{1}Joint Quantum Institute, National Institute of Standards and Technology and the University of Maryland, College Park, Maryland 20742, USA\\
\authormark{2}Department of Physics, American University, Washington, D.C. 20016, USA\\
\authormark{3}Department of Physics, Williams College, Williamstown, Massachusetts 01267, USA\\
\authormark{4}Quantum Measurement Division, National Institute of Standards and Technology, Gaithersburg, Maryland 20899, USA}

\email{\authormark{*}paul.lett@nist.gov} %% email address is required



\begin{abstract}
Homodyne detection is often used for interferometers based on nonlinear optical gain media. For the configuration of a seeded, ``truncated SU(1,1)'' interferometer Anderson et al.~\cite{PhysRevA.95.063843} showed how to optimize the homodyne detection scheme and demonstrated theoretically that it can saturate the quantum Cramer-Rao bound for phase estimation. In this work we extend those results by taking into account loss in the truncated SU(1,1) interferometer and determining the optimized homodyne detection scheme for phase measurement. Further, we build a truncated SU(1,1) interferometer and experimentally demonstrate that this optimized scheme achieves a reduction in noise level, corresponding to an enhanced potential phase sensitivity, compared to a typical homodyne detection scheme for a two-mode squeezed state. In doing so, we also demonstrate an improvement in the degree to which we can beat the standard quantum limit with this device. 
\\*

\end{abstract}

 % REPLACE WITH CORRECT OCIS CODES FOR YOUR ARTICLE, MINIMUM OF TWO; Avoid using the OCIS codes for “General” or “General science” whenever possible.
%For a complete list of OCIS codes, visit: https://www.osapublishing.org/oe/submit/ocis/

%%%%%%%%%%%%%%%%%%%%%%% References %%%%%%%%%%%%%%%%%%%%%%%%%

%%%%%%%%%%%%%%%%%%%%%%%%%%  body  %%%%%%%%%%%%%%%%%%%%%%%%%%
\section{Introduction}
Interferometers allow for extremely sensitive measurements of optical phase. The optimum sensitivity of a specific interferometer is related to the optical quantum state that is used for sensing the phase and the placement of the phase object in the interferometer. Coherent light with an average of $n$ photons in a detection interval gives an uncertainty in an ideal interferometric phase measurement equal to  $1/\sqrt{n}$, known as the standard quantum limit (SQL). The sensitivity of interferometers can be improved beyond the SQL by using quantum states of light instead of, or in addition to, coherent states. Such improvements have been demonstrated by injecting squeezed light into an interferometer~\cite{PhysRevD.23.1693, PhysRevLett.59.278,PhysRevLett.59.2153} and with the use of Fock or other non-classical states~\cite{PhysRevLett.71.1355, Nature429.161.164, NaturePhotonics5.43.47}. In another class of interferometers, instead of injecting the quantum light into the interferometer, the non-classical state of light is generated inside the interferometer. An SU(1,1) interferometer suggested by Yurke et al.~\cite{PhysRevA.33.4033} and the variation constructed here both fall into the latter class.

An SU(1,1) interferometer is formed by replacing the beam splitters in a Mach-Zehnder interferometer with two nonlinear optical (NLO) media~\cite{PhysRevA.33.4033}, as shown in Fig.~\ref{su(1,1)_interferometer}. In an SU(1,1) interferometer, there are two input states. In the case depicted here, one input is a coherent beam of amplitude $\alpha$ (with mean photon number $|\alpha|^2$), and the other input is vacuum $|0\rangle$. The nonlinear interaction of the coherent seed beam with a strong pump beam in the medium amplifies the coherent beam to produce a probe beam with mean photon number $G|\alpha|^2$ and also produces correlated light with mean photon number $(G-1)|\alpha|^2$ in another mode known as the conjugate, where $G$ is the gain of the nonlinear process. In the present work, we use a 4-wave mixing (4WM) process in $^{\text{85}}\text{Rb}$ vapor~\cite{Boyer544, OptLett32178180} to produce the probe and the conjugate beams. Individually, the probe and the conjugate beams are noisier than coherent states with the same amplitude but together they form a pair of beams that constitute a two-mode squeezed state of light. The seeded mode, i.e., the probe, contains the phase object within the interferometer. After the probe has passed through the phase object, which puts a phase shift $\delta\phi$ on the probe, the two modes interact in a second 4WM process.
The interferometer output depends on the sum of the phases of the probe and the conjugate beams relative to the phase of the pump beam at the second nonlinear medium as well as the small phase shift $\delta\phi$. The relative phases of the probe and the conjugate beams are locked with respect to the pump beam in the interferometer such that the sensitivity to the measurement of $\delta\phi$ is optimized. The outputs from the SU(1,1) interferometer can be measured either by direct detection or by homodyne detection. Processing of the electrical signals from the outputs gives the value of the small phase shift $\delta\phi$. In our experimental setup using 4WM in $^{\text{85}}\text{Rb}$ vapor, one of the advantages of a homodyne detection is that by selecting the appropriate frequency modes using local oscillators it removes any extra pump beam scattering that could induce excess noise in the measurement. 

\begin{figure}[tb!]
  \centering
    \begin{subfigure}{1\textwidth}
  \centering
    \includegraphics[width=0.6\textwidth]{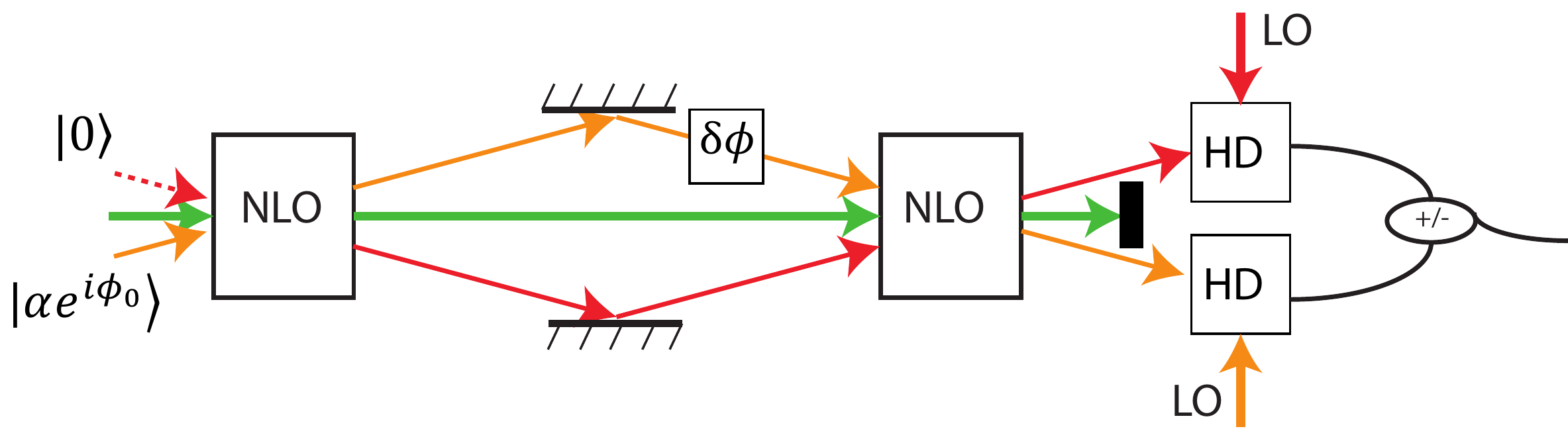}

\caption{}\label{su(1,1)_interferometer}

\end{subfigure}%

\begin{subfigure}{0.5\textwidth}
  \centering
    \includegraphics[width=1\textwidth]{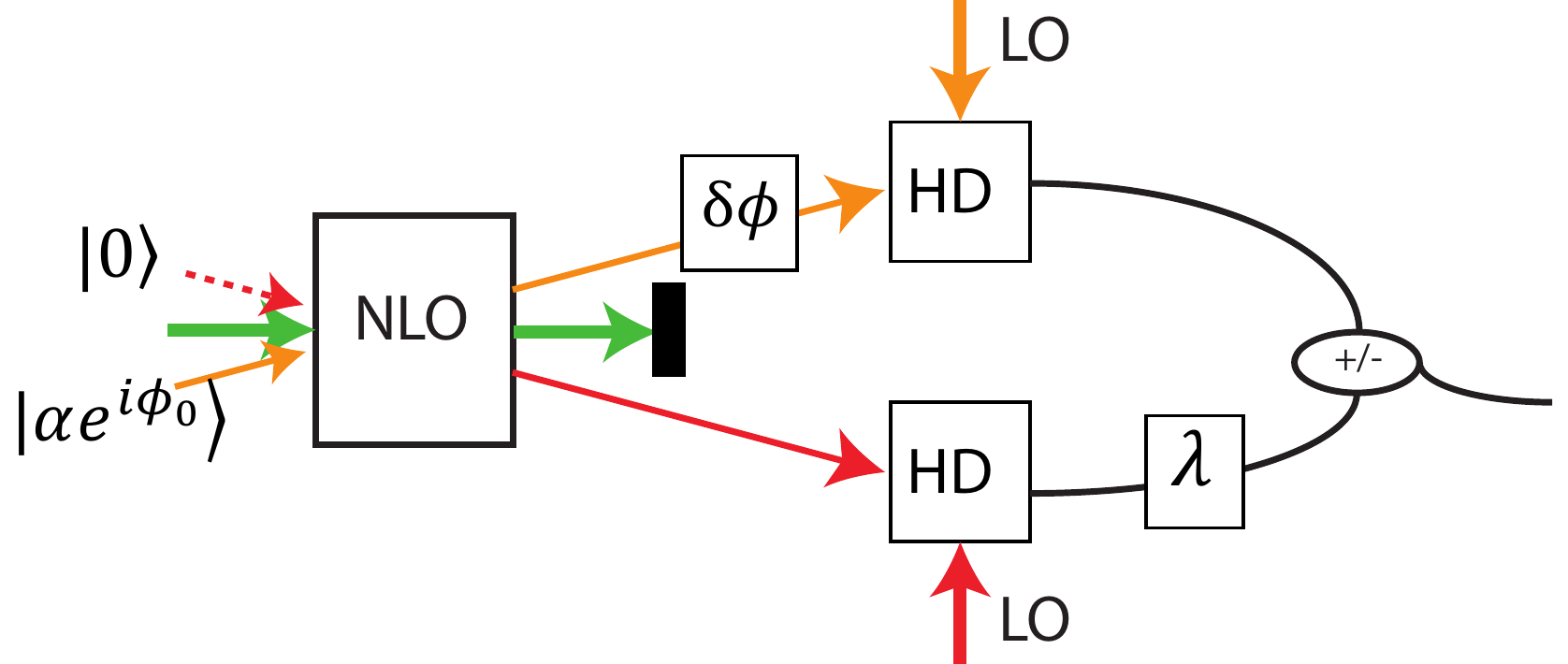}

\caption{}\label{truncated_su(1,1)_interferometer}
\end{subfigure}%
\begin{subfigure}{0.5\textwidth}
  \centering
    \includegraphics[width=0.75\textwidth]{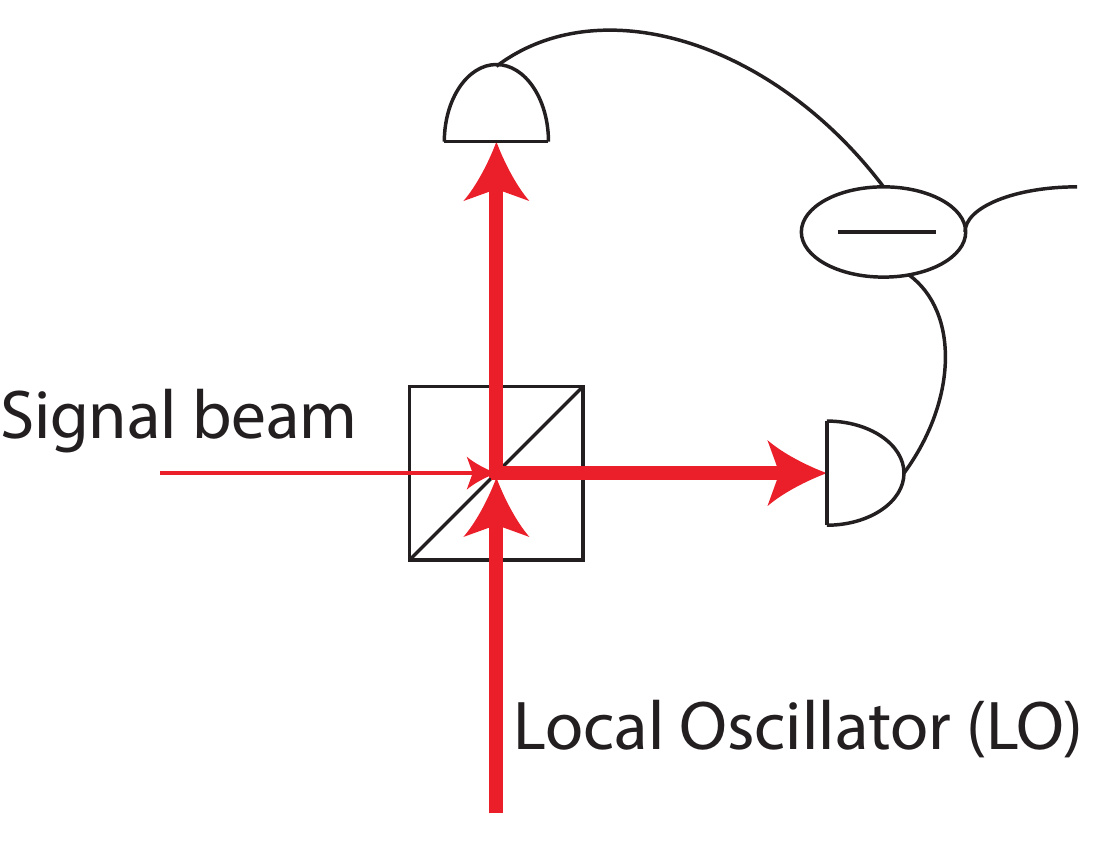}

\caption{}\label{homodyne_measurement}
\end{subfigure}%

\caption{Schematics of SU(1,1) type interferometers. A coherent state $|\alpha e^{i\phi_0}\rangle$ with amplitude $\alpha$ and phase $\phi_{0}$ and a vacuum state $|0\rangle$ are mixed with a strong pump beam in an NLO medium to produce a probe beam and a conjugate beam, which together form a two-mode squeezed state. A phase shift $\delta\phi$ is applied to the seeded arm, i.e., the probe beam. a) The SU(1,1) interferometer recombines the probe and the conjugate beams in another nonlinear medium. The diagram shows homodyne detectors (HD) after the second cell that measure the output and thus the phase shift $\delta\phi$. One could also perform direct intensity detection on the outputs to measure the phase shift $\delta\phi$. b) The truncated SU(1,1) interferometer sends the probe and the conjugate beams directly to homodyne detectors~\cite{PhysRevA.95.063843, Anderson:17}. The output of the conjugate beam homodyne detector is attenuated by a factor $\lambda$ using an electronic attenuator before it is combined with the output of the probe homodyne detector to perform the phase measurement. We discuss the electronic attenuation in Sec. 2. c) A schematic of a homodyne detector for phase measurement~\cite{hudelist_kong_liu_jing_ou_zhang_2014, PhysRevA.85.023815, NaturePhotonics9.577.581, Science337.1514.1517} in which a signal beam interferes with a strong local oscillator (LO) of the same spatial mode and frequency as the signal beam on a 50:50 beam splitter. The difference in the photo-currents of the two outputs gives the quadrature of the signal beam amplified by the amplitude of the LO. }\label{interferometer_image}
\end{figure}

Theoretically, an SU(1,1) interferometer should have sub-SQL phase sensitivity that scales as $1/n$ instead of $1/\sqrt{n}$, where $n$ is the average number of photons in a detection interval. This possibility of enhanced quantum metrology has attracted the attention of many authors to this device~\cite{PhysRevA.33.4033, NewJ.Phys.12.083014, PhysRevA.86.023844, hudelist_kong_liu_jing_ou_zhang_2014, arXiv.1705.02662v2, PhysRevA.85.023815}. In our previous works~\cite{PhysRevA.95.063843, Anderson:17}, we presented a simplified variation of the SU(1,1) interferometer, a ``truncated SU(1,1) interferometer,'' which removes the second nonlinear process and sends the probe and conjugate beams directly to the homodyne detectors, as shown in Fig.~\ref{truncated_su(1,1)_interferometer}. In these works, we showed that this setup can achieve the same phase sensitivity as that of an SU(1,1) interferometer~\cite{PhysRevA.95.063843, Anderson:17}.

An important metric for evaluating the sensitivity of an interferometer is the quantum Fisher information ($\mathscr{F}_{Q}$), which depends only on the quantum state of the light inside the interferometer and the phase shift on the quantum state and is independent of the detection scheme. The quantum Cramer Rao bound (QCRB)~\cite{Helstrom, Paris} relates the the minimum detectable phase shift $\Delta\phi$ to the quantum Fisher information by $(\Delta\phi)^2\geq1/\mathscr{F}_{Q}$. The sensitivity of a device depends on the chosen detection scheme and is limited by the QCRB for an ideal measurement~\cite{Sparaciari:15, PhysRevLett.102.040403, PhysRevA.93.023810}. The losses present in the device further limit the sensitivity of the measurement from achieving the QCRB ~\cite{PhysRevLett.102.040403, PhysRevA.86.023844, PhysRevA.57.4004, PhysRevA.69.043616, PhysRevA.75.053805, 0953-4075-40-14-001, Gilbert:08, PhysRevA.81.033819}. Enough loss on the quantum state prevents the interferometer from performing any better than the SQL.

Anderson et al.~\cite{PhysRevA.95.063843} have shown that a lossless truncated SU(1,1) interferometer approaches the QCRB for phase sensitivity, asymptotically, for large quantum correlations (i.e., a high squeezing level) between the probe and the conjugate beams. At small correlations (i.e., a low squeezing level), the phase sensitivity of an equally weighted joint quadrature measurement differs significantly from the QCRB, which makes it a non-optimal measurement~\cite{PhysRevA.95.063843}. Anderson et al.~\cite{PhysRevA.95.063843} showed theoretically that by changing the relative weights of the signals from the two homodyne detectors (i.e., by changing $\lambda$ in Fig.~\ref{truncated_su(1,1)_interferometer}), one can saturate the QCRB with any amount of correlations between the probe and conjugate, i.e., squeezing, which is set by the gain ($G$) of the 4WM process. In this manuscript, we extend their analysis to account for the loss in the interferometer particularly in the low gain regime, and we examine the improvement in phase sensitivity, both theoretically and experimentally, offered by these optimized measurements over the equally weighted joint quadrature measurement for a truncated SU(1,1) interferometer. We show theoretically that the optimized phase measurement scheme depends on both the gain and the loss in the system. We construct a truncated SU(1,1) interferometer and show that by using this modified, weighted joint homodyne detection scheme, one can reduce noise in the phase measurement and hence can achieve enhanced phase sensitivities compared to the equally weighted scheme.

\section{Background}
Given a measured observable $\hat{M}$ depending on a phase parameter $\phi$, the sensitivity of the phase measurement can be evaluated using the signal-to-noise ratio (SNR) defined by
\begin{equation}\label{eq:signal_to_noise_ratio}
\text{SNR}=\frac{(\partial_{\phi}M)^2 (\delta\phi)^2}{\Delta^2 M},
\end{equation}
where $\partial_{\phi}M$ is the partial derivative of the observable $\hat{M}$ with respect to $\phi$, $\Delta^2M$ is the noise in the measurement of $\hat{M}$ and $\delta\phi$ is a small deviation in $\phi$ being measured. In our interferometer, the phase $\phi$ defines the operating point of the interferometer and depends on the relative phases of the probe and conjugate beams and their local oscillators, respectively. To maximize the SNR for a measurement of $\delta\phi$ the phase $\phi$ is locked to the most sensitive point. The minimum detectable phase ($\delta\phi_{\text{min}}$) from the measurement can be obtained by setting the SNR=1, which gives the phase sensitivity of the measurement $\Delta\phi=\delta\phi_{\text{min}}=\left(\frac{\Delta^2 M}{\partial_{\phi}M}\right)^{1/2}$.

For a truncated SU(1,1) interferometer, the optimal sensitivity is achieved when each homodyne detector is locked to measure the phase quadrature of its input beams~\cite{PhysRevA.95.063843, Anderson:17}. The sum of the two locked homodyne detector outputs is the joint phase sum quadrature defined by
\begin{equation}\label{eq:joint_phase_sum_operator}
\hat{M}_{Q}=(\hat{X}_{p}+\hat{X}_{c}),
\end{equation}
where $\hat{X}_{p}$ and $\hat{X}_{c}$ represent the phase quadratures of the probe and the conjugate beams, respectively. In our setup, the joint phase quadrature of the probe and the conjugate is the squeezed quadrature for the two-mode squeezed light with noise lower than the joint quadrature noise of two independent vacuum states. One can view the truncated SU(1,1) interferometer in Fig.~\ref{truncated_su(1,1)_interferometer} as being composed of a simple interferometer involving the signal (probe) beam and the LO used in its homodyne detector. This arm provides the entire signal from the device. The signal depends on the phase difference between the probe and its local oscillator as well as the small phase shift, $\delta\phi$, being applied by the phase object. The phase $\phi$ is locked to measure the phase quadrature of both the probe and the conjugate beams, where it maximizes the SNR for the phase shift $\delta\phi$ being measured. The conjugate arm of the device is then used to subtract off correlated noise that can be detected independently in this arm of the device. The degree of correlation between the noise in the two arms determines the resulting improvement in the SNR~\cite{PhysRevA.95.063843, Anderson:17}.

\begin{figure}[tb!]
  \centering
    \includegraphics[width=0.45\textwidth]{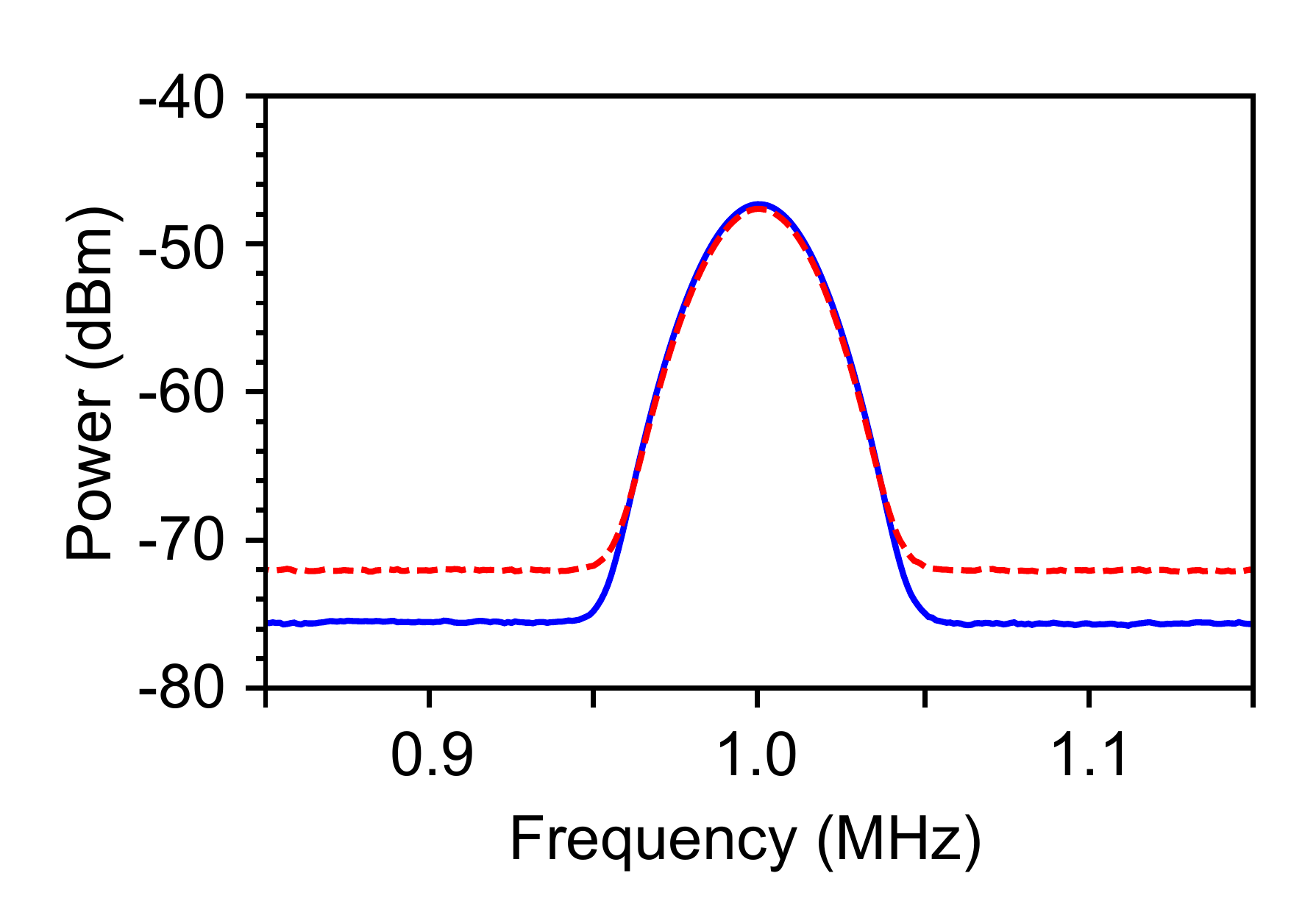}

\caption{Measured power on a spectrum analyzer as a function of frequency in a truncated SU(1,1) interferometer using two-mode squeezed light (blue, solid line) and with a coherent beam (red, dashed line) of the same optical power in the phase sensing (probe beam) arm of the interferometer, as shown in Ref.~\cite{Anderson:17}. The coherent beam SNR represents the SQL, and the difference in the noise floor of the traces shows a 4.0(1) dB improvement in SNR with the squeezed light over the SQL. The 4WM gain used for the measurement was 3.3, higher than what we discuss here.}\label{snr_data}
\end{figure}

In our previous experimental work~\cite{Anderson:17}, we showed an improvement in phase sensitivity beyond the SQL using a truncated SU(1,1) interferometer. In that work, we passed the local oscillator for the probe beam through an optical phase modulator, which modulated the beam at 1 MHz and produced a signal for our measurement. When measuring the joint phase quadrature of the squeezed light, we observed the signal shown as the solid blue curve in Fig.~\ref{snr_data}. Measuring the joint phase quadrature of coherent beams of the same power produced the red dashed curve in Fig.~\ref{snr_data}, which represents the SQL. The overlapping peaks show the same power in the signal due to the phase modulation, but the baseline noise level associated with the squeezed light was reduced due to the correlations between the probe and conjugate beams. The difference in the SNR of these two signals indicates a 4.0(1) dB improvement in phase sensitivity beyond the SQL. 

\begin{figure}[tb!]
  \centering
    \includegraphics[width=0.45\textwidth]{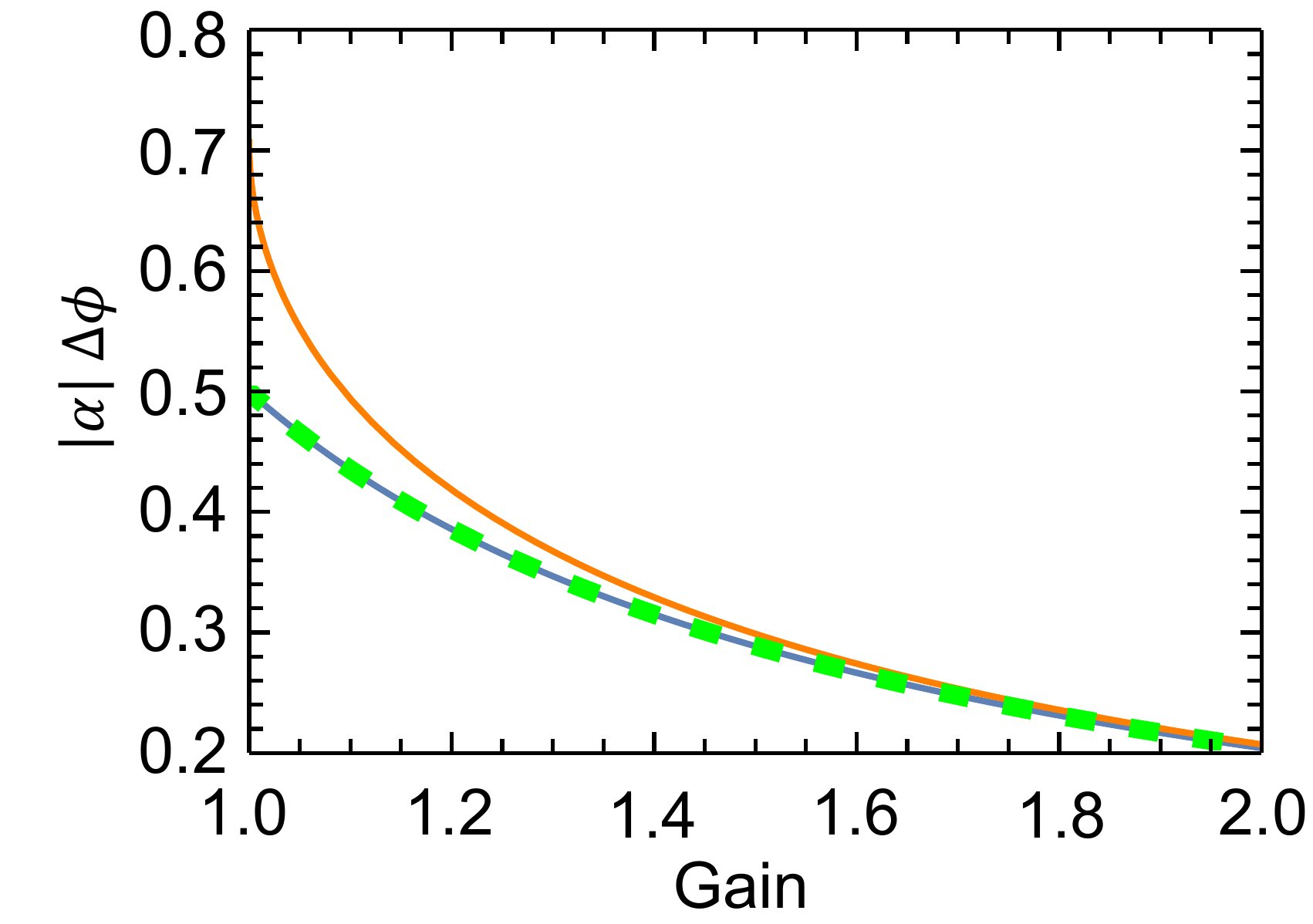}

\caption{Theoretical peak sensitivity, multiplied by the amplitude ($|\alpha|$) of the coherent seed beam of the interferometer, achieved by an ideal lossless truncated SU(1,1) interferometer as a function of gain in the 4WM process. The solid orange curve shows the phase sensitivity of the observable $\hat{M}_{Q}$ and the solid blue curve represents the phase sensitivity of the observable $\hat{M}_{\lambda_{opt} Q}$ (as defined in the text). The thick dashed green curve indicates the QCRB for the two-mode squeezed state.}\label{sensitivity_comparision_with_qfi}
\end{figure}

As discussed earlier, for an ideal lossless truncated SU(1,1) interferometer, measurement of the balanced observable $\hat{M}_Q$ from Eq.~(\ref{eq:joint_phase_sum_operator}) saturates the QCRB only asymptotically with increasing correlations between the probe and conjugate (increasing gain of the 4WM process). For small 4WM gain, there is a significant difference between the sensitivity achieved with the measurement of $\hat{M}_Q$ and the QCRB as shown in Fig.~\ref{sensitivity_comparision_with_qfi}. Anderson et al.~\cite{PhysRevA.95.063843} have briefly discussed this deviation and proposed the following measurement for saturating the QCRB:
\begin{equation}\label{eq:operator}
\hat{M}_{\lambda Q}=\hat{X}_{p}+\lambda \hat{X}_c,
\end{equation}
where $\lambda$ is a scalar with a value between 0 and 1. $\hat{M}_{\lambda Q}$ approaches $\hat{M}_Q$ as $\lambda$ approaches 1. We refer to $\hat{M}_{\lambda Q}$ as the weighted operator and $\hat{M}_Q$ as the balanced operator. Depending on the gain of the 4WM process and the losses in the interferometer, there is an optimum value of $\lambda, \lambda_{opt}$, at which the operator $\hat{M}_{\lambda Q}$ optimizes the phase sensitivity. We define the operator $\hat{M}_{\lambda_{opt}Q}$ as $\hat{M}_{\lambda Q}$ with $\lambda=\lambda_{opt}$, and we call it the optimized weighted operator. For a lossless truncated SU(1,1) interferometer $\lambda_{opt}=\text{tanh}(2r)$, where $r$ is related to the gain $(G)$ of the 4WM process via $r=\text{cosh}^{-1}\sqrt{G}$~\cite{PhysRevA.95.063843}. In Fig.~\ref{sensitivity_comparision_with_qfi}, we plot the theoretical sensitivity for a lossless truncated SU(1,1) interferometer with the measurement of $\hat{M}_{\lambda_{opt}Q}$ as a function of the gain $G$, which saturates the QCRB for all the values of 4WM gain. Similarly, for a full SU(1,1) interferometer, using our theoretical analysis we find that the QCRB can be saturated at small gains of the first nonlinear medium with the measurement of $\hat{M}_{\lambda_{opt}Q}$ but that the values of $\lambda_{opt}$ are different than the ones for the truncated SU(1,1) interferometer and depend on the gain of the second nonlinear process. In this manuscript, we will discuss the optimal measurements only for a truncated SU(1,1) interferometer.

Experimentally, $\hat{M}_{\lambda Q}$ can be measured by attenuating the output of the conjugate homodyne detector and then adding it to the output of the probe homodyne detector. We note that changing the attenuation $\lambda$ on the conjugate homodyne detector does not modify the signal from the probe homodyne detector. Thus, adjusting the attenuation $\lambda$ only changes the noise power and does not affect the signal due to the phase object placed in the probe homodyne detector. Hence, to observe the change in SNR as we modify the attenuation $\lambda$, it is sufficient to monitor the noise power of the joint homodyne detection. Therefore, in the remainder of this paper, we will quantify improvement in SNR only in terms of noise power variation.

\begin{figure}[tb!]
  \centering

\begin{subfigure}{0.45\textwidth}
  \centering
    \includegraphics[width=1\textwidth]{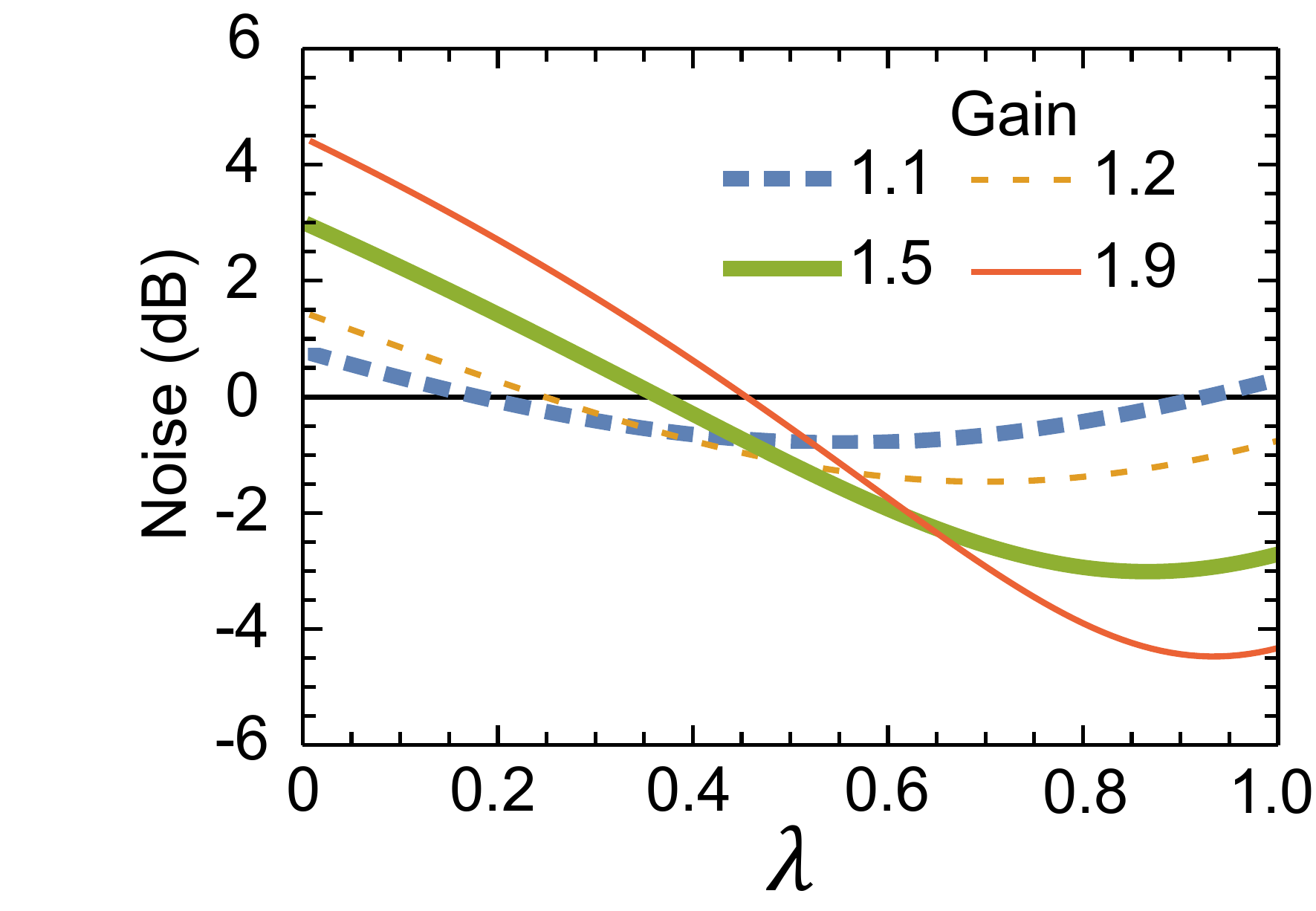}
\caption{}\label{noise_vs_eta}
\end{subfigure}%
\hfill
\begin{subfigure}{0.45\textwidth}
  \centering
    \includegraphics[width=1\textwidth]{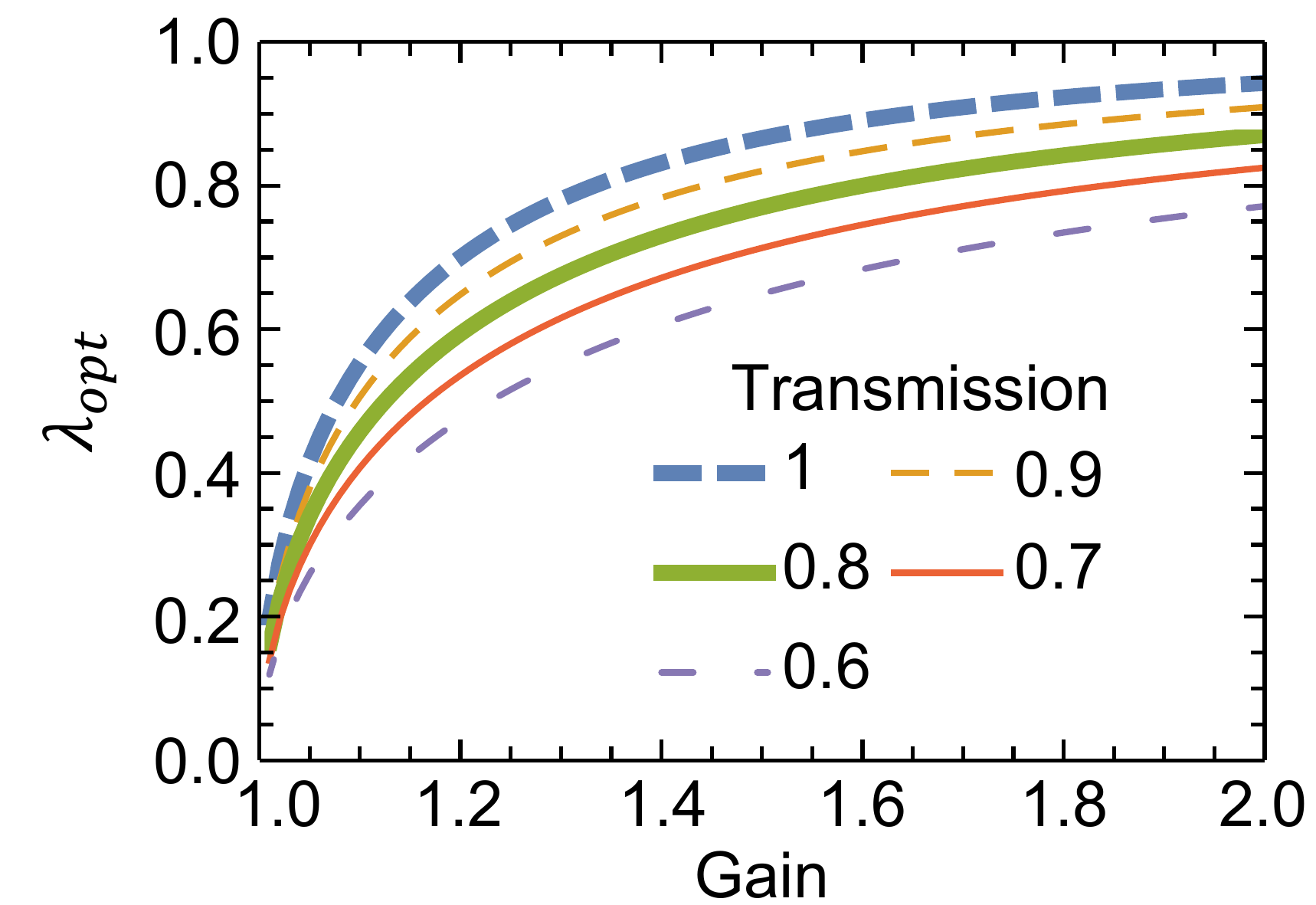}

\caption{}\label{optimumcA_with_gain_loss}
\end{subfigure}%

\caption{a) Theoretical noise in the measurement of the operator $\hat{M}_{\lambda Q}$ in an ideal lossless truncated SU(1,1) interferometer as a function of $\lambda$. The zero on the vertical axis represents the shot noise of a single homodyne detector without any attenuation on its output, calculated by replacing the probe or the conjugate beam with vacuum. b) The value of $\lambda_{opt}$ as a function of the 4WM gain at different optical transmissions of the probe ($\eta_{p}$) and the conjugate beams ($\eta_{c}$) in the interferometer, assuming $\eta_{p}=\eta_{c}$.}
\end{figure}

We plot the variation of the theoretical noise power of $\hat{M}_{\lambda Q}$ as a function of the attenuation $\lambda$ for a lossless truncated SU(1,1) interferometer at different values of 4WM gain in Fig.~\ref{noise_vs_eta}. The minimum in the noise power shows the point of optimal phase sensitivity and corresponds to the value of $\lambda=\lambda_{opt}$. The thick dashed blue (uppermost) curve in Fig.~\ref{optimumcA_with_gain_loss} shows the variation of $\lambda_{opt}$ as a function of gain of the 4WM process in a lossless truncated SU(1,1) interferometer and indicates that $\lambda_{opt}$ approaches 1 ($\hat{M}_{\lambda_{opt} Q}$ approaches $\hat{M}_Q$) as the gain increases.

Thus far, we have discussed only lossless interferometers, but all real-world systems have some amount of loss in them. The loss in the interferometer comes from multiple sources: the optical components used in building the experiment, the detection loss due to inefficient photodiodes used to detect light, and the electronic noise power of the electrical circuits of the detection system, for example. Any loss on the probe and conjugate beams decreases the correlations between them and thus changes the value of $\lambda_{opt}$ that optimizes the phase sensitivity of the measurement. In a lossy interferometer, we find 
\begin{equation}\label{eq:optimum_lambda_parameter}
\lambda_{opt}=\frac{\sqrt{\eta_{p}}\sqrt{\eta_c} \text{sinh}(2r)}{1-\eta_c+\eta_c \text{cosh}(2r)},
\end{equation}
where $\eta_p$ and $\eta_c$ are the transmissions of the probe and the conjugate beams in the interferometer. The losses on the probe and the conjugate beams are given by 1$-\eta_p$ and 1$-\eta_c$, respectively. The above expression reduces to $\lambda_{opt}=\text{tanh} (2r)$ for a lossless interferometer with $\eta_p = \eta_c=1$. We plot the variations of $\lambda_{opt}$ as a function of 4WM gain for several values of the probe and the conjugate beams' transmissions in Fig.~\ref{optimumcA_with_gain_loss}. The presence of loss on the probe and the conjugate beams lowers the value of $\lambda_{opt}$ that optimizes the sensitivity and slows down its saturation to 1 as the gain of the 4WM increases, which is a consequence of reduced correlations between the probe and the conjugate beams due to the optical loss. 

Before proceeding further, we can try to understand the improvement in phase sensitivity of the measurement using the weighted operator $\hat{M}_{\lambda_{opt}Q}$ over the balanced operator $\hat{M}_{Q}$. For a gain of 1, i.e., when the 4WM process is switched off, we can replace the probe and the conjugate beams with a coherent beam and vacuum, which are uncorrelated with each other. In this case, the conjugate homodyne detector only adds excess noise when added to the probe homodyne detector and hence removing the conjugate detector gives the best phase measurement, i.e., $\lambda_{opt}=0$ is the optimum parameter value for measuring phase. For large 4WM gains, the probe and conjugate beams are highly correlated, and hence the noise reduction achieved with $\lambda_{opt}=1$ (i.e., measurement $\hat{M}_Q$) gives the best phase sensitivity. For the intermediate values of the 4WM gain, we observe the value of $\lambda_{opt}$ between 0 and 1.

The improvement in phase sensitivity using the optimized weighted operator $\hat{M}_{\lambda_{opt}Q}$ over the balanced operator $\hat{M}_Q$ may also suggest a further improvement over the SQL. SQL is defined by the best phase sensitivity of a Mach-Zehnder interferometer. If $n$ is the number of photons in a detection interval passing through the phase object placed in one of the arms of the interferometer then the SQL is given by $(\Delta\phi_{\text{SQL}})^2=1/2n$. Here, we will suggest two possible definitions of the SQL and analyze the potential improvement in phase sensitivity relative to each one. First, we make an analogy of the truncated SU(1,1) interferometer to a Mach-Zehnder interferometer~\cite{Anderson:17}; this definition of SQL (SQL1) uses two homodyne detectors. Later, we give a more stringent definition of SQL (SQL2) that uses only one homodyne detector.

\begin{figure}[tb!]
  \centering
    \includegraphics[width=0.45\textwidth]{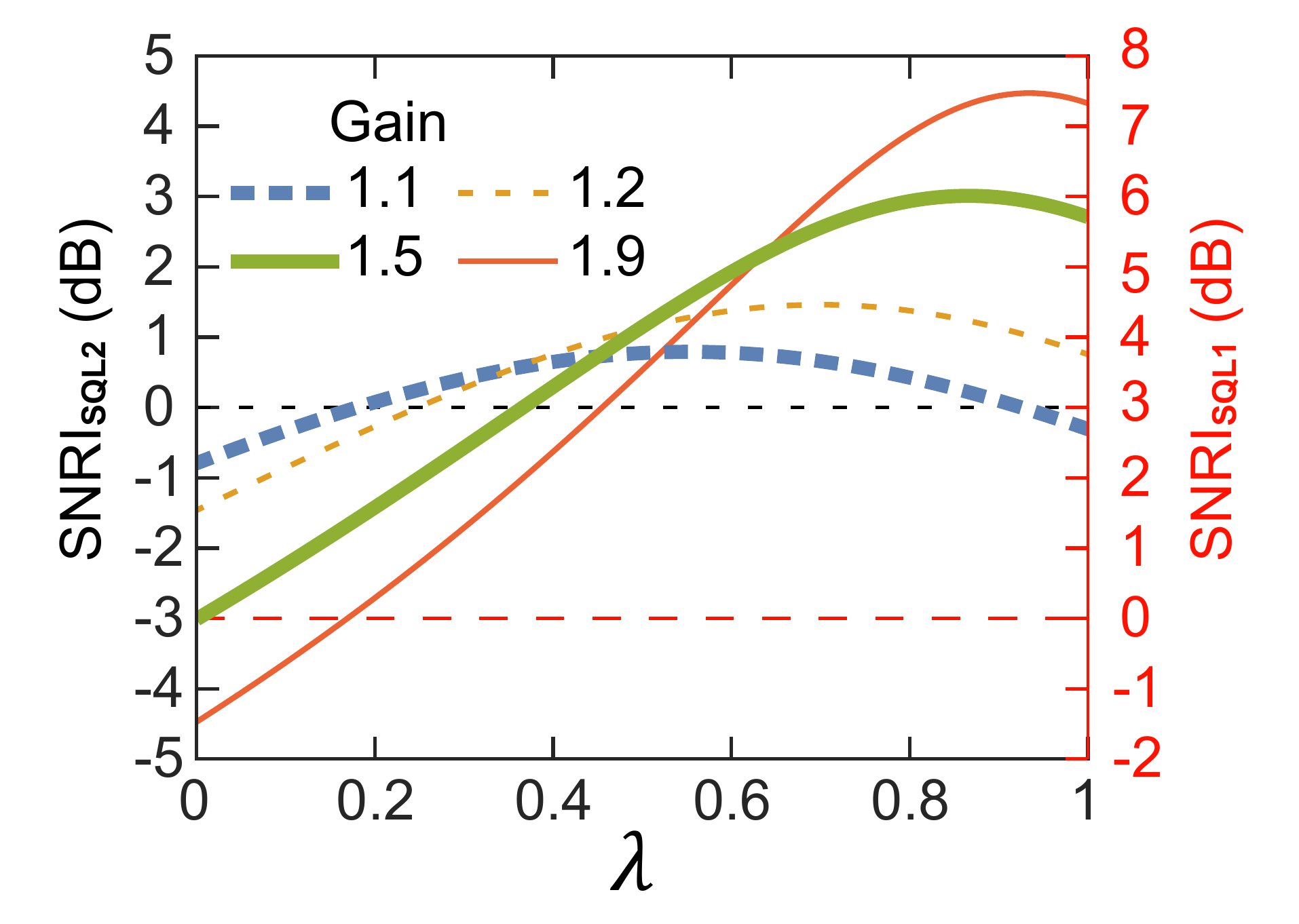}

\caption{Theoretical improvement in the SNR as a function of $\lambda$ with the measurement $\hat{M}_{\lambda Q}$ over the conditions of SQL2 ($\text{SNRI}_{\text{SQL2}}$), and SQL1 ($\text{SNRI}_{\text{SQL1}}$) for a lossless truncated SU(1,1) interferometer. The measurement conditions are defined in the text and the different curves are for different values of the 4WM gain. Positive values represent improved SNR (and phase sensitivity) over the SQL as defined for those conditions.}\label{theoretical_improvement_over_SQL1_SQL2}
\end{figure}

In our first possible definition of the SQL, we compare the sensitivity of our truncated SU(1,1) interferometer with the sensitivity of a truncated Mach-Zehnder interferometer in which the second beam splitter is replaced with two homodyne detectors. We add the outputs of the two homodyne detectors to measure the operator $\hat{M}_Q$, i.e., $\hat{M}_{\lambda Q}$ with $\lambda=1$ for measuring the SQL. At its optimum operating point the truncated configuration of a Mach-Zehnder interferometer has the same sensitivity as the best sensitivity of a standard Mach-Zehnder interferometer~\cite{Anderson:17} and hence can be used as a measure of SQL. We send the same number of coherent  beam photons through the phase object in the truncated Mach-Zehnder interferometer as are there in the probe beam of the truncated SU(1,1) interferometer which passes through the phase object. We refer to the sensitivity of this configuration as SQL1 and represent it by $\Delta\phi_{SQL1}$. In Fig.~\ref{theoretical_improvement_over_SQL1_SQL2}, on the right vertical axis, we plot the improvement in SNR (SNRI) of a truncated SU(1,1) interferometer over the SNR associated with SQL1, given by $\text{SNRI}_{\text{SQL1}}$. The SNRI associated with a given SQL (SQLi, with i=1,2) is given by 
\begin{equation}\label{eq:snri_over_sql1}
\text{SNRI}_{\text{SQLi}}=\text{SNR}_{\text{tSUI}}-\text{SNR}_{\text{SQLi}}=10~\text{Log}_{10}\left[\left(\frac{\Delta\phi_{\text{SQLi}}}{\Delta\phi_{\text{tSUI}}}\right)^2\right],
\end{equation}
where $\text{SNR}_{\text{tSUI}}$ and $\Delta\phi_{\text{tSUI}}$ represents the SNR and the sensitivity of the truncated SU(1,1) interferometer with the measurement of $\hat{M}_{\lambda Q}$, respectively. We derive this equation by relating the phase sensitivity to the SNR using Eq.~(\ref{eq:signal_to_noise_ratio}) and then taking a difference between the SNR associated with SQLi and the SNR of the truncated SU(1,1) interferometer.

In our experimental setup, one could argue that SQL1 does not represent the optimum coherent beam sensitivity and that this measurement scheme can be improved to give a better phase estimate. When we replace the second beam splitter of a Mach Zehnder with two homodyne detectors for the two coherent beams, then the homodyne detector on the beam which does not pass through the phase object only adds excess noise. Hence removing that homodyne detector could reduce the noise and thus improve the SNR of the measurement. To make measurements using this scheme, we only perform homodyne detection on the beam passing through the phase object and call the limiting phase sensitivity of this configuration SQL2, represented by $\Delta\phi_{SQL2}$. On the left vertical axis of Fig.~\ref{theoretical_improvement_over_SQL1_SQL2}, we show the SNRI of a truncated SU(1,1) interferometer with the measurement of the weighted operator $\hat{M}_{\lambda Q}$ over the SNR associated with SQL2, i.e., $\text{SNRI}_\text{SQL2}$, as a function of the attenuation $\lambda$. Here the improvement optimizes for $\lambda=\lambda_{opt}$. Since SQL2 is a more stringent definition of the SQL, the improvement over SQL2 is not as large as the improvement over SQL1. The difference in the improvement over each SQL, $\text{SNRI}_{\text{SQL1}}-\text{SNRI}_{\text{SQL2}}$, is 3 dB, which arises due to the absence of the second homodyne detector in SQL2.

We also note from Fig.~\ref{theoretical_improvement_over_SQL1_SQL2} that the phase sensitivity of the equally weighted measurement $\hat{M}_Q$ at small gains (thick dashed blue curve, gain=1.1) does not beat the SQL2 whereas the weighted measurement $\hat{M}_{\lambda_{opt}Q}$ does. This further highlights the importance of using the measurement $\hat{M}_{\lambda_{opt}Q}$ instead of $\hat{M}_Q$ in the low gain regime. 

\section{Experimental demonstration}
To demonstrate a reduction in noise of the interferometer and hence a potential improvement in phase sensitivity using the measurement $\hat{M}_{\lambda_{opt} Q}$ we have constructed a truncated SU(1,1) interferometer~\cite{Anderson:17} using 4WM in hot $^{85}\text{Rb}$ vapor~\cite{Boyer544}. We perform homodyne detection on both the probe and the conjugate beams to measure the joint quadrature. The output of each homodyne detector is split into two parts: the low frequency (DC) and high frequency (AC) parts. The DC parts are used to phase lock the homodyne detectors to measure the phase quadrature of the beams. The AC parts are used for the joint quadrature measurement. To measure $\hat{M}_{\lambda Q}$, we electronically attenuate the AC output of the conjugate homodyne detector and then combine it with the AC output of the probe homodyne detector. The combined signal then goes to a spectrum analyzer, where we measure the noise power of the joint homodyne detector. We measure the noise at a frequency of 1~MHz within a resolution bandwidth of 100 kHz.

\begin{figure}[tb!]
\centering
\begin{subfigure}[t]{1\textwidth}
  \centering
  \includegraphics[width=0.45\linewidth,clip=false]{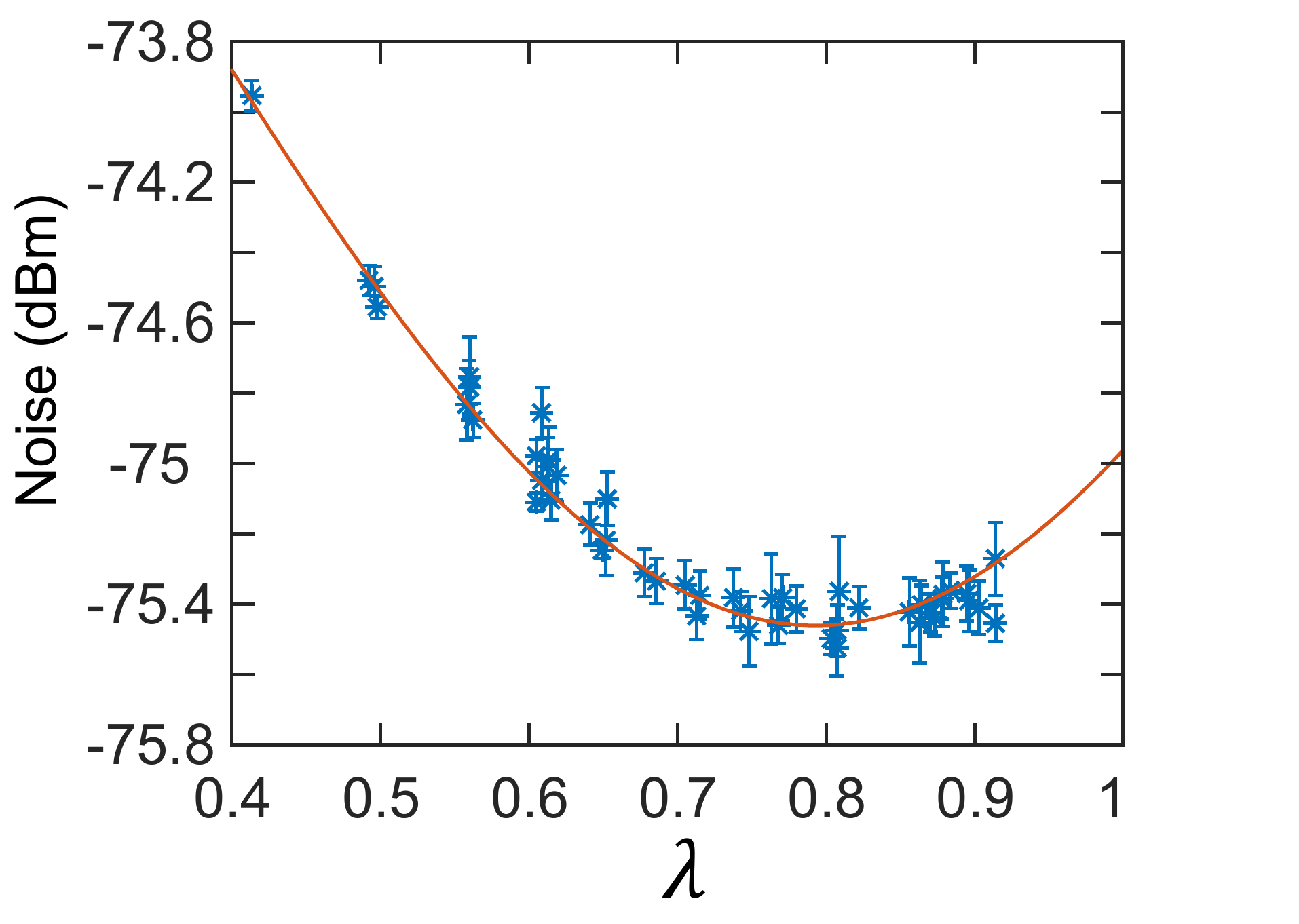}
\hfill
  \includegraphics[width=0.455\linewidth,clip=false]{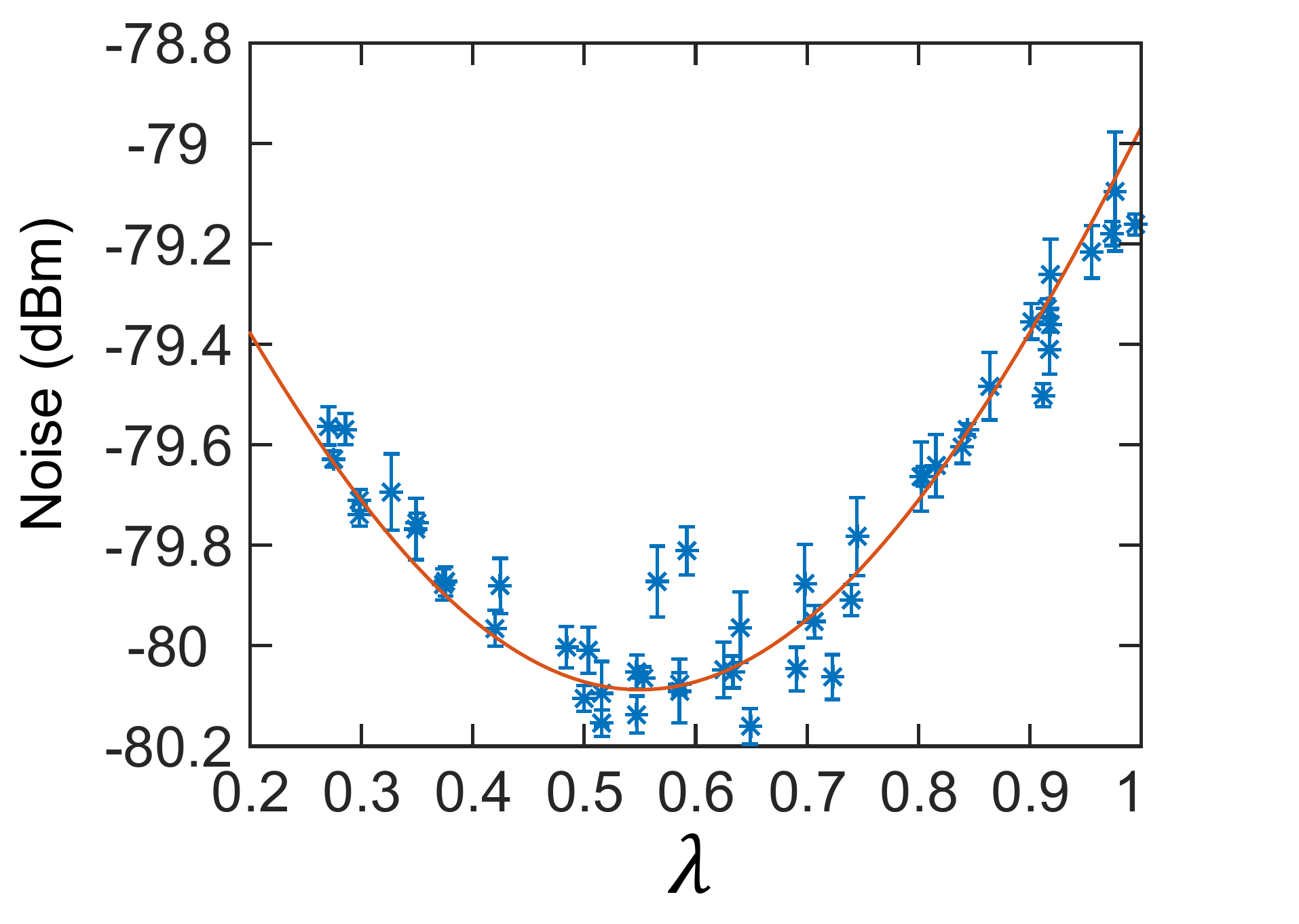}
  \caption{}
  \label{single_seed_thing_for_LO_differemce_test_noise}
\end{subfigure}
\begin{subfigure}[t]{1\textwidth}
  \centering
  \includegraphics[width=0.45\linewidth]{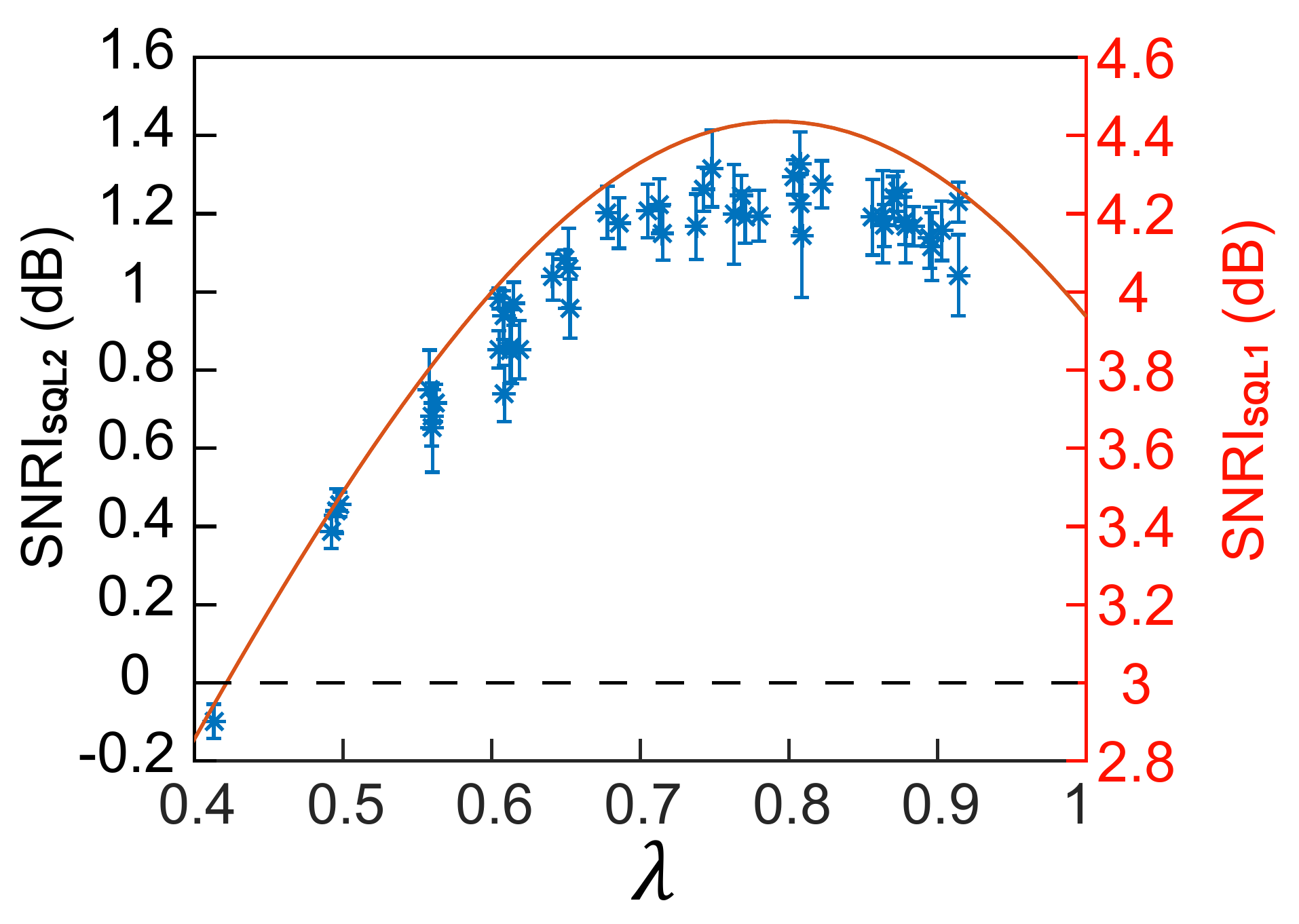}
\hfill
  \includegraphics[width=0.45\linewidth]{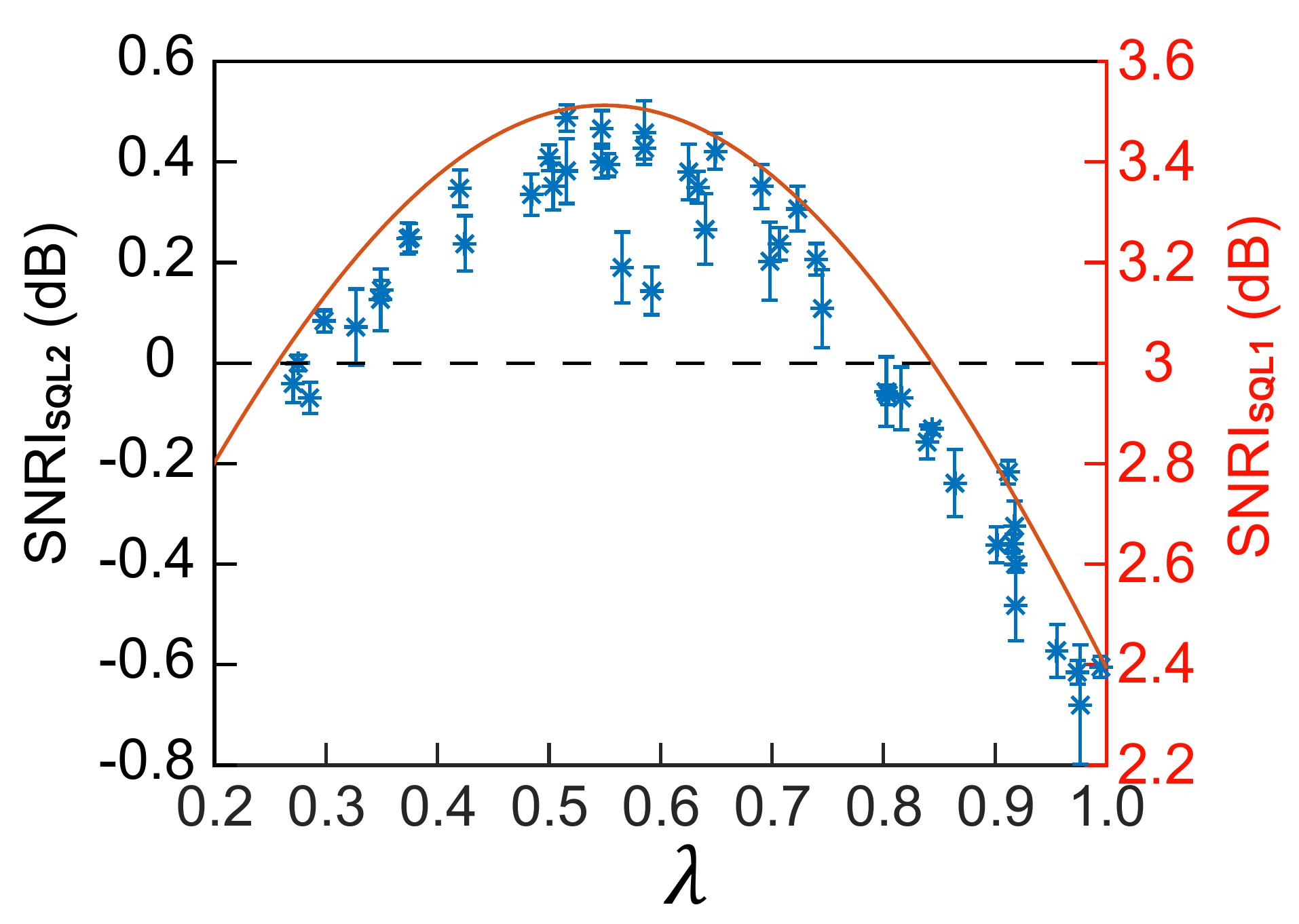}
  \caption{}
  \label{single_seed_thing_for_LO_differemce_test_improvement_over_SQL2SQL1}
\end{subfigure}
\hfill
\caption{a) Noise in the joint homodyne detection measuring $\hat{M}_{\lambda Q}$ with squeezed light as a function of the attenuation parameter $\lambda$. b) Improvement in the SNR as a function of the attenuation $\lambda$ with the measurements $\hat{M}_{\lambda Q}$ over the conditions of SQL1 ($\text{SNRI}_{\text{SQL1}}$) and SQL2 ($\text{SNRI}_{\text{SQL2}}$). The left side plots have an estimated 4WM gain of 1.67, a probe transmission of 76$\%$ and a conjugate transmission of 79$\%$. The right side plots have an estimated 4WM gain of 1.2, a probe transmission of 73$\%$ and a conjugate transmission of 76$\%$. The gain and the loss values were estimated from the theoretical fit of the data. In these fits, we assume that the probe suffers 3$\%$ more loss than the conjugate beam, which we measure experimentally and occurs because the probe is closer to the absorption resonance of $^{\text{85}}\text{Rb}$.}
\label{Squeezed noise floor, imoprovent over SQL2 and improvememnt over SQL1}
\end{figure}

We measure the noise power of $\hat{M}_{\lambda Q}$ as a function of $\lambda$ at different values of the 4WM gain, two of which are shown in Fig.~\ref{single_seed_thing_for_LO_differemce_test_noise}. The presence of noise minima corresponds to the point $\lambda_{opt}$, where the sensitivity optimizes. We also measure $\text{SNRI}_\text{SQL1}$ and $\text{SNRI}_\text{SQL2}$ as a function of $\lambda$ in Fig.~\ref{single_seed_thing_for_LO_differemce_test_improvement_over_SQL2SQL1}. We measure the SQL1 and SQL2 by turning off the 4WM process (by blocking the pump beam) and making the coherent seed equal in power to the probe beam produced by the 4WM process. We then perform the appropriate measurements using the homodyne detectors based on the measurements of SQL1 and SQL2, as described in Sec. 2. The plots in Fig.~\ref{Squeezed noise floor, imoprovent over SQL2 and improvememnt over SQL1} show a trend similar to that in the theoretical simulations of Fig.~\ref{noise_vs_eta} and~\ref{theoretical_improvement_over_SQL1_SQL2}. We fit the noise curve in Fig.~\ref{single_seed_thing_for_LO_differemce_test_noise} with the theory using the 4WM gain, the loss on the probe and the conjugate, and a scaling parameter to adjust to the absolute value of the noise. The scaling parameter takes into account any DC shift in the plot because of the power of the local oscillators or the imperfect locking of the homodyne detectors. The 4WM gain and the losses on the individual beams define the shape of the noise curve as a function of $\lambda$ and the position of the minimum, i.e., $\lambda_{opt}$ in the plot. The fitted parameters, i.e., the gain and the losses on the probe and the conjugate, were then used to put theoretical curves on $\text{SNRI}_\text{SQL1}$ and $\text{SNRI}_\text{SQL2}$. 

The theoretical curve in Fig.~\ref{single_seed_thing_for_LO_differemce_test_improvement_over_SQL2SQL1} passes through the upper range of the uncertainties on the data points. This behavior is attributed to the imperfect locking of the phases of the probe and the conjugate homodyne detectors. In our setup, $\hat{M}_{\lambda Q}$, for a given $\lambda$, represents the measurement of the squeezed quadrature for our two-mode squeezed light. Any instability in phase locking gives excess noise in the joint homodyne detection and hence reduces the improvement in SNR. The theoretical curve does not take into account the imperfection in phase locking and represents the maximum achievable improvement in the system for the given gain and loss in the interferometer. Hence the theoretical curve passes through the upper range of the uncertainties on the data points, showing the maximum improvement possible. 

Figure~\ref{single_seed_thing_for_LO_differemce_test_improvement_over_SQL2SQL1} (left vertical axis) also demonstrates that at small 4WM gains, the measurement $\hat{M}_{\lambda_{opt}Q}$ can improve the phase sensitivity beyond SQL2 whereas the measurement of the balanced operator $\hat{M}_Q$, i.e., $\hat{M}_{\lambda Q}$ with $\lambda=1$, does not give any enhanced sensitivity, as discussed in Sec. 2.

\begin{figure}[tb!]
  \centering
    \includegraphics[width=0.45\textwidth]{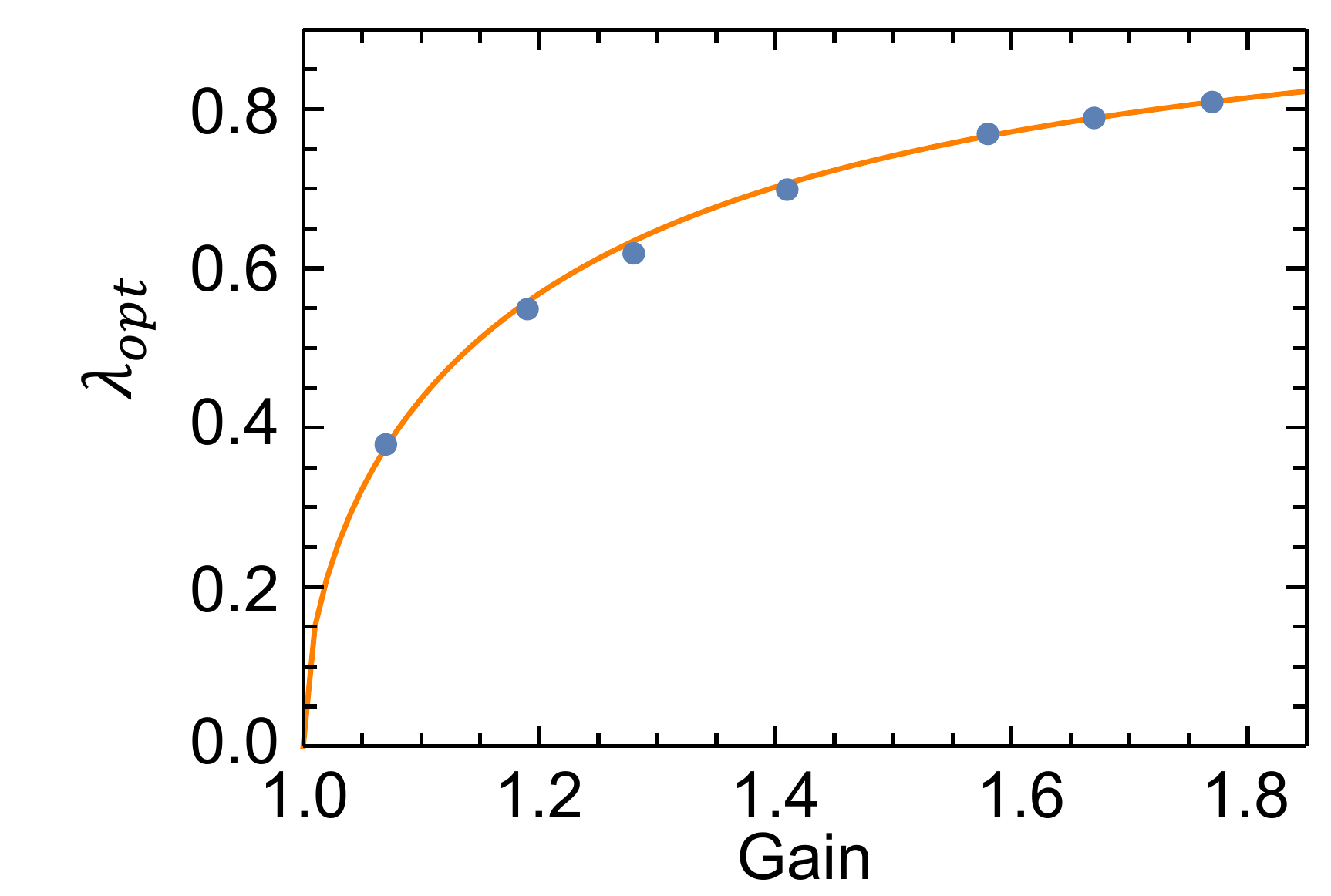}

\caption{$\lambda_{opt}$ as a function of the 4WM gain. The points are experimental measurements determined from plots like those in Figs.~\ref{single_seed_thing_for_LO_differemce_test_noise} and~\ref{single_seed_thing_for_LO_differemce_test_improvement_over_SQL2SQL1}. A theoretical curve is generated using a probe beam transmission of 74.5$\%$ and a conjugate beam transmission of 77.5$\%$. These values represent the typical losses in our system.}\label{data_optimumcA_with_gain_loss}
\end{figure}

We also estimate $\lambda_{opt}$ values for different 4WM gains by extracting the minima from the data in Fig.~\ref{single_seed_thing_for_LO_differemce_test_noise}, as well as from similar data for other 4WM gains, which we show in Fig.~\ref{data_optimumcA_with_gain_loss}. The experimental data points fit the theory (from Fig.~\ref{optimumcA_with_gain_loss}) quite well and hence Fig.~\ref{data_optimumcA_with_gain_loss} serves as an experimental verification of the trend shown in Fig.~\ref{optimumcA_with_gain_loss}. 

\section{Conclusion}
The class of interferometers based on active media has become of interest in the pursuit of metrological devices that can beat the standard quantum limit.  The optimized measurements for these devices are not the same under all conditions; here we have investigated this optimization for the truncated SU(1,1) interferometer under a range of gain and loss conditions. We have experimentally confirmed the improvement in SNR, and thus phase sensitivity, with the measurement of the weighted operator $\hat{M}_{\lambda_{opt}Q}$ over the usual balanced operator $\hat{M}_Q$. We have shown an improvement in phase sensitivity over the SQL with the measurement $\hat{M}_{\lambda_{opt}Q}$ under the very conservative conditions for setting the SQL (SQL2, as defined above) and experimentally demonstrated its utility at small 4WM gains, when the measurement $\hat{M}_Q$ does not produce any improvement over the SQL2 limit. Lastly, we have verified the trend in $\lambda_{opt}$ values as a function of 4WM gain using the experimental data. While the optimized measurement that saturates the QCRB will not be the same for other interferometer geometries or under other conditions, the present work demonstrates a procedure to optimize homodyne measurements that should be generally useful.
\hfill \break
\hfill \break
\textbf{Funding: }National Science Foundation (NSF); Air Force Office of Scientific Research (AFOSR).	

\bibliography{LO_difference_paper}
\bibstyle{plain}
\end{document}